\documentclass[trackchanges,twocolumn]{aastex701}

\newcommand{\kms}{\mathrm{km\,s}^{-1}}

\usepackage{amsmath}	
\usepackage{amssymb}	

\begin{document}

\title{Discovery of an Extremely Luminous Type II Cepheid in the Andromeda Giant Stellar Stream: Evidence for a Hierarchical Triple with an Inner Binary Merger}

\author[orcid=0009-0007-5623-2475]{Pinjian Chen}
\affiliation{National Astronomical Observatories, Chinese Academy of Sciences, Beijing 100101, P.\,R.\,China}
\affiliation{School of Astronomy and Space Science, University of the Chinese Academy of Sciences, Beijing, 100049, P.\,R.\,China}
\email{} 

\author[orcid=0000-0003-2472-4903]{Bingqiu Chen}
\affiliation{South-Western Institute for Astronomy Research, Yunnan University, Kunming, Yunnan 650091, P.\,R.\,China}
\email[show]{bchen@ynu.edu.cn} 

\author[orcid=0000-0001-7084-0484]{Xiaodian Chen}
\affiliation{National Astronomical Observatories, Chinese Academy of Sciences, Beijing 100101, P.\,R.\,China}
\affiliation{School of Astronomy and Space Science, University of the Chinese Academy of Sciences, Beijing, 100049, P.\,R.\,China}
\affiliation{Institute for Frontiers in Astronomy and Astrophysics, Beijing Normal University, Beijing 102206, P.\,R.\,China}
\email[show]{chenxiaodian@nao.cas.cn} 

\author[orcid=0000-0003-2471-2363]{Haibo Yuan}
\affiliation{School of Physics and Astronomy, Beijing Normal University, Beijing 100875, P.\,R.\,China}
\affiliation{Institute for Frontiers in Astronomy and Astrophysics, Beijing Normal University, Beijing 102206, P.\,R.\,China}
\email{} 

\author[orcid=0000-0002-0349-7839]{Jianrong Shi}
\affiliation{National Astronomical Observatories, Chinese Academy of Sciences, Beijing 100101, P.\,R.\,China}
\affiliation{School of Astronomy and Space Science, University of the Chinese Academy of Sciences, Beijing, 100049, P.\,R.\,China}
\affiliation{School of Physics and Technology, Nantong University, Nantong 226019, P.\,R.\,China}
\email{} 

\author[orcid=0000-0003-4489-9794]{Shu Wang}
\affiliation{National Astronomical Observatories, Chinese Academy of Sciences, Beijing 100101, P.\,R.\,China}
\affiliation{School of Astronomy and Space Science, University of the Chinese Academy of Sciences, Beijing, 100049, P.\,R.\,China}
\email{}

\author[orcid=0000-0002-6647-3957]{Chunqian Li}
\affiliation{National Astronomical Observatories, Chinese Academy of Sciences, Beijing 100101, P.\,R.\,China}
\affiliation{School of Astronomy and Space Science, University of the Chinese Academy of Sciences, Beijing, 100049, P.\,R.\,China}
\email{}

\author[orcid=0009-0002-1155-284X]{Jiyu Wang}
\affiliation{National Astronomical Observatories, Chinese Academy of Sciences, Beijing 100101, P.\,R.\,China}
\affiliation{School of Astronomy and Space Science, University of the Chinese Academy of Sciences, Beijing, 100049, P.\,R.\,China}
\email{}

\author[orcid=0009-0001-5015-0387]{Jianxing Zhang}
\affiliation{National Astronomical Observatories, Chinese Academy of Sciences, Beijing 100101, P.\,R.\,China}
\affiliation{School of Astronomy and Space Science, University of the Chinese Academy of Sciences, Beijing, 100049, P.\,R.\,China}
\email{}

\author[orcid=0000-0003-1218-8699]{Yi Ren}
\affiliation{Department of Astronomy, College of Physics and Electronic Engineering, Qilu Normal University, Jinan 250200, P.\,R.\,China}
\email{} 
 
\begin{abstract}
We report the discovery of LAMOST J0041+3948, the most luminous post-AGB Type II Cepheid (TIIC) known, located in the Andromeda Giant Stellar Stream. Its spectral energy distribution (SED) exhibits a strong near-infrared excess, indicating the presence of a circumbinary dusty disk and hence binarity. SED fitting yields an effective temperature of $T_{\rm eff}=6738_{-262}^{+234}$\,K and a post-AGB luminosity of $\log(L/L_{\odot})=4.32_{-0.08}^{+0.07}$. Comparison with theoretical evolutionary tracks suggests a $\sim2.0-4.0\,M_{\odot}$ progenitor when accounting for a possible scattered-light contribution. ZTF Light curves reveal a pulsation period of 89\,d that lies close to the period-luminosity relation for long-period RV Tauri stars. Follow-up spectroscopy reveals clear $s$-process enrichment and signatures consistent with an accretion disk around the companion. The inferred progenitor is significantly younger and more massive than a typical stream member, suggesting that an additional mechanism such as a stellar merger is required. We propose a formation channel in which the present post-AGB binary descends from a hierarchical triple system. In this scenario, the inner binary merged after the system was displaced to its current location by the galaxy merger event, and the resulting massive merger remnant subsequently evolved into the extremely luminous post-AGB star observed today.

\end{abstract}

\section{Introduction} \label{sec:intro}
Post-asymptotic giant-branch (post-AGB) represents a short-lived ($\sim 10^3-10^5$\,yr) but crucial phase in the late evolution of low-to-intermediate-mass stars \citep{VanWinckel2003}. After shedding most of their envelope mass, the central remnant contracts and evolves toward higher effective temperatures while retaining nearly constant luminosity \citep[e.g.][]{Miller2016}. When such stars cross the instability strip in the Hertzsprung-Russell (HR) diagram, they pulsate as Type II Cepheids (TIICs). The post-AGB TIICs occupy the high-luminosity end of the Population II Cepheid family, typically exhibiting long pulsation periods\footnote{Note that throughout this paper, the pulsation period refers to the fundamental period for TIICs.} ($>$20\,d) and are often categorized as RV Tauri stars \citep[RVTs;][]{Alcock1998, Wallerstein2002, Soszynski2008, Bono2020}. Although the evolutionary channels of post-AGB stars are not yet fully understood, a consensus has been reached in recent years that binarity plays an important, and possibly decisive, role in shaping these systems \citep[for comprehensive reviews, see][]{VanWinckel2018, VanWinckel2025}.

Binary post-AGB stars are characterized by a distinctive ``disk-type" spectral energy distribution (SED), exhibiting a broad near-infrared (IR) excess that signals hot dust near the sublimation temperature. It is now well established that this SED morphology indicates the presence of a stable, compact (circumbinary) disk of gas and dust \citep[e.g.,][]{deRuyter2006, Gezer2015, Hillen2016, Hillen2017, VanWinckel2017, Kluska2022,Corporaal2023}, and has been repeatedly validated as a robust indicator of binarity through radial velocity monitoring \citep[e.g.,][]{vanWinckel2009, Manick2017, Oomen2018}. These systems are typically wide binaries with orbital periods of $\sim100-3000$\,d and frequently exhibit eccentric orbits \citep{Oomen2018}.

Ongoing interaction processes occur in post-AGB binary systems, with a striking observational fingerprint in the form of chemical anomalies called ``depletion" in their photospheres \citep[e.g.][]{VanWinckel2003, Maas2005}: refractory elements are underabundant, whereas volatiles maintain near-original abundances. A plausible explanation is gas-dust separation in the circumbinary disk, followed by the re-accretion of metal-poor gas onto the post-AGB star's surface \citep{Waters1992}. Consistent with this picture, all known depleted sources exhibit disk-type SEDs \citep{Gezer2015}. Additionally, phase-dependent absorption features in H$\alpha$ profile reveal the presence of outflows or jets, likely launched from the accretion disk surrounding the companion \citep{Bollen2017, Bollen2019}. The circumstellar disk serves as the main reservoir for accretion, maintaining a high accretion rate \citep{Bollen2020, Bollen2022} and strengthening the connection between accretion and depletion patterns in these systems.

In this Letter we report the discovery of a peculiar binary post-AGB TIIC in M31's halo, LAMOST J004146.97+394817.0 (hereafter J0041+3948). This object was originally noticed during our effort to compile a comprehensive catalog of M31 sources from the LAMOST database \citep{Chen2025a}. Its large negative radial velocity, negative parallax, and small proper motions all favor membership in M31 rather than being a foreground Milky Way star. Although it lies well outside M31's optical disk, its brightness implies a high intrinsic luminosity, motivating follow-up investigation. We summarize its observed and derived properties in Table~\ref{tab: Properties}. 

Throughout this paper, we adopt a distance to M31 of $785 \pm 25$\,kpc \citep{McConnachie2012}, corresponding to a linear scale of approximately 13.7\,kpc\,deg$^{-1}$. The optical radius is taken as $R_{25} = 95.3\arcmin$ \citep{deVaucouleurs1991}. We assume the galaxy disk is viewed with an inclination of 77\arcdeg\ and a position angle of 38\arcdeg\ \citep{Walterbos1987}.

\section{Observations} \label{sec:observation}

\subsection{Spectroscopy}
J0041+3948 was first observed by the Large Sky Area Multi-Object Fiber Spectroscopic Telescope \citep[LAMOST;][]{Cui2012} in its low-resolution mode ($R\sim 1800$) on 12 December 2014, and the data were processed using the standard LAMOST pipeline \citep{Luo2015, Xiang2015}. We carried out follow-up spectroscopy on 14 September 2024 and 12 October 2024 using the Double Spectrograph \citep[DBSP;][]{Oke1982} mounted on the 200-inch Hale Telescope (P200) at Palomar Observatory. The observations employed the D55 dichroic, with the 1200/5000 grating on the blue arm and the 600/10000 grating on the red arm. A 1.5\arcsec\ slit was consistently used across all exposures to match the seeing conditions. This setup yielded a spectral resolution (FWHM) of 2.11\,\AA\ in the blue ($\sim3800-5400$\,\AA) and 4.19\,\AA\ in the red ($\sim5700-9100$\,\AA). A summary of the spectroscopic observations is provided in Appendix~\ref{sec:spec_log}. The DBSP data were reduced with the \texttt{PypeIt} package \citep{pypeit1,pypeit2} following the standard procedures.

\begin{table}[t]
\begin{center}
\setlength{\tabcolsep}{5.pt}
\caption{\label{tab: Properties}Observed and Derived Physical Properties of J0041+3948}
\begin{tabular}{lr}
\hline
\hline
Properties$^a$ & Value \\
\hline
$\alpha,\delta$ (J2000) & 00:41:46.97, +39:48:17.04\\
$\,R_{\text{proj}}$ (deg) & $1.48$ \\
$\varpi$ (mas), RUWE & $-0.035 \pm 0.146$, 0.97\\
$\mu_{\alpha}$, $\mu_{\delta}$ (mas\,yr$^{-1}$) & $-0.214 \pm 0.135$, $0.304 \pm 0.113$\\
$G$, $G_{\text{BP}}-G_{\text{RP}}$ (mag) & $18.366\, \pm\, 0.002$, $0.535\, \pm\, 0.033$\\
Radial Velocity ($\kms$) & $-511.2\pm1.6$\\
Pulsation Period (days) & 89\\
\hline
\textbf{Stellar$^b$} & Value\\
$T_{\text{eff}}$ (K) & $6738_{-262}^{+234}$ \\
$\log g$ (dex) & $1.8_{-0.8}^{+0.7}$ \\
$R/R_{\odot}$ & $106_{-7}^{+8}$ \\
$E(B-V)$ (mag) & $0.05_{-0.04}^{+0.04}$ \\
Distance Modulus (mag) &  $24.41_{-0.14}^{+0.15}$\\
$\log L_{\star}/L_{\odot}$ & $4.32_{-0.08}^{+0.07}$ \\

\hline
\textbf{Disk$^b$} & Value \\
$T_{\text{dust}}$ (K) & $957_{-50}^{+48}$\\
$\log L_{\text{dust}}/L_{\odot}$ & $3.82_{-0.06}^{+0.06}$ \\
\hline
\end{tabular}
\end{center}
\tablenotetext{a}{Astrometric, photometric, and color information are adopted from Gaia DR3.} 
\tablenotetext{b}{Uncertainties correspond to $2\sigma$ confidence intervals from the SED fit.}
\end{table}

\begin{figure*}
\centering 
\includegraphics[width=1.0\textwidth]{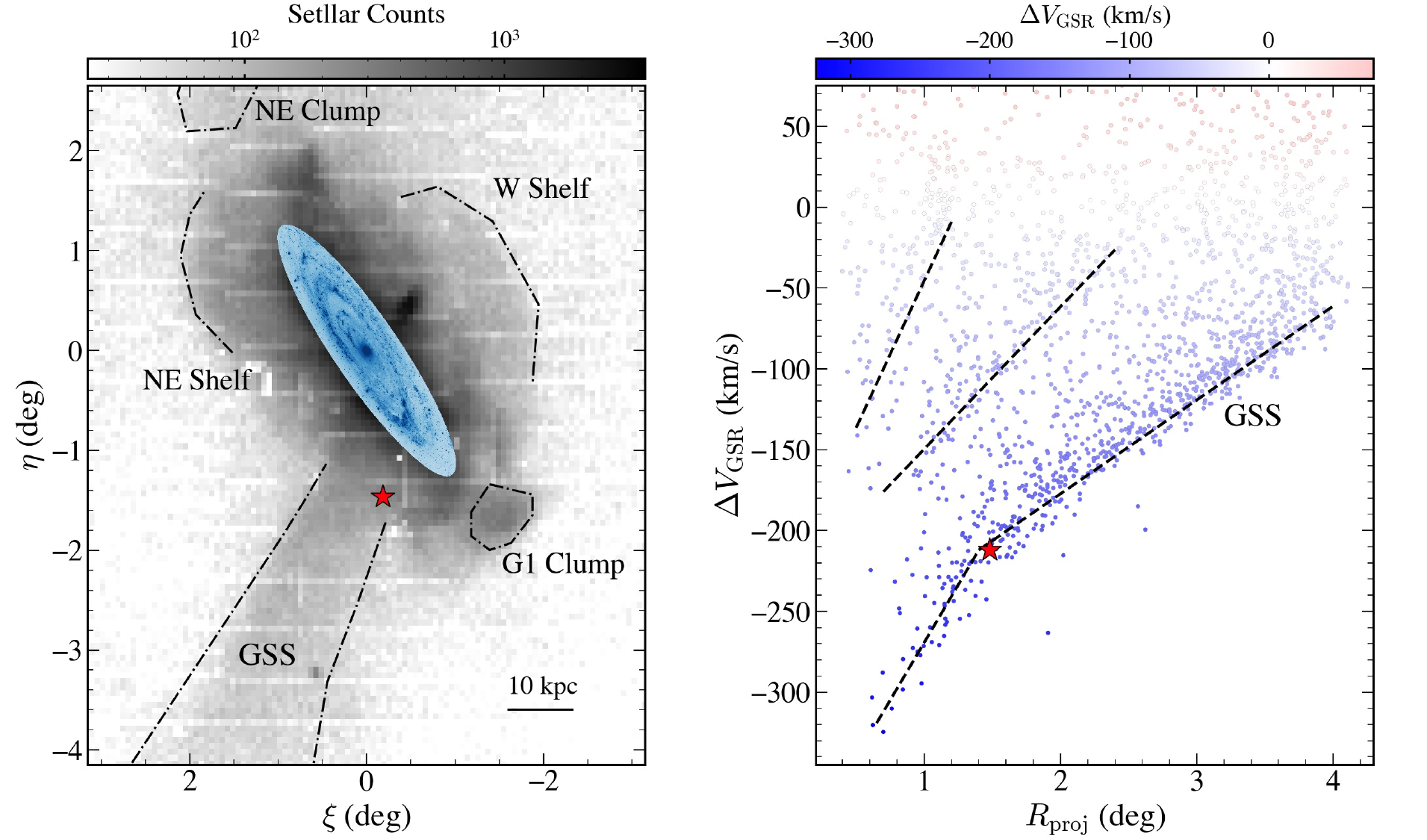}
\caption{
Left: Location of J0041+3948 (large red asterisk) in M31-centric coordinates overlaid on a PAndAS stellar density map (logarithmic scale). The M31 disk region within $R_{25}$ is indicated using a GALEX NUV image. Major substructures are outlined and labeled. Right: Position-velocity diagram. J0041+3948 is marked by the large red asterisk, and the selected DESI stars from \citet{Dey2023} are color coded by their velocities. Black dashed lines outline the three main kinematic features identified by \citet{Dey2023}, with the GSS component labeled.
\label{fig: location}}
\end{figure*}

\subsection{Photometry} 
We compiled archival photometry to construct the multiwavelength SED of J0041+3948. We adopted the Sloan Digital Sky Survey (SDSS) DR16 \citep[$ugriz$;][]{sdssdr16}, Pan-STARRS1 DR2 \citep[$grizy$;][]{PS1} and \textit{Gaia} DR3 \citep[$GG_{\text{BP}}G_{\text{RP}}$;][]{gaiadr3} in the optical, and CatWISE2020 \citep[$W1W2$;][]{catwise} for the near-IR. The \textit{WISE} images show that our target is partially blended with a nearby foreground star, and thus the measured photometry may be affected. Additionally, we performed $JHK$-band photometry on UKIRT/WFCAM images obtained in November 2006 \citep{ukirt}, following the method of \citet{Ren2021}. Zwicky Transient Facility \citep[ZTF;][]{Bellm2019, Masci2019} DR23 light curves within 1'' were retrieved from the NASA/IPAC Infrared Science Archive \citep[IRSA;][]{ipac} to assess variability. 

\section{Association with the GSS} \label{sec:association}

J0041+3948 resides about $1.5\arcdeg$ southeast of M31's center, well outside the optical disk (see the left panel of Figure\,\ref{fig: location}). To visualize the stellar halo, we used the Pan-Andromeda Archaeological Survey \citep[PAndAS;][]{McConnachie2009, McConnachie2018} catalog and selected red giant branch (RGB) stars with the polygonal color-magnitude cut in ($g-i$, $g$) defined by \citet{Dey2023}. As shown in the figure, M31's halo is rich in tidal features (e.g. clumps,  shelves, and streams), with the most prominent overdensity being the Giant Stellar Stream \citep[GSS;][]{Ibata2001, Ibata2004}. This massive, extended stream is thought to be a tidal debris originating from a past galaxy merger. 
J0041+3948 projects onto the region where the GSS intersects the inner stellar halo, corresponding to $\sim20$\,kpc in the plane of the sky and $\sim49$\,kpc in the plane of the disk.

\begin{figure*}
\centering 
\includegraphics[width=1.05\textwidth]{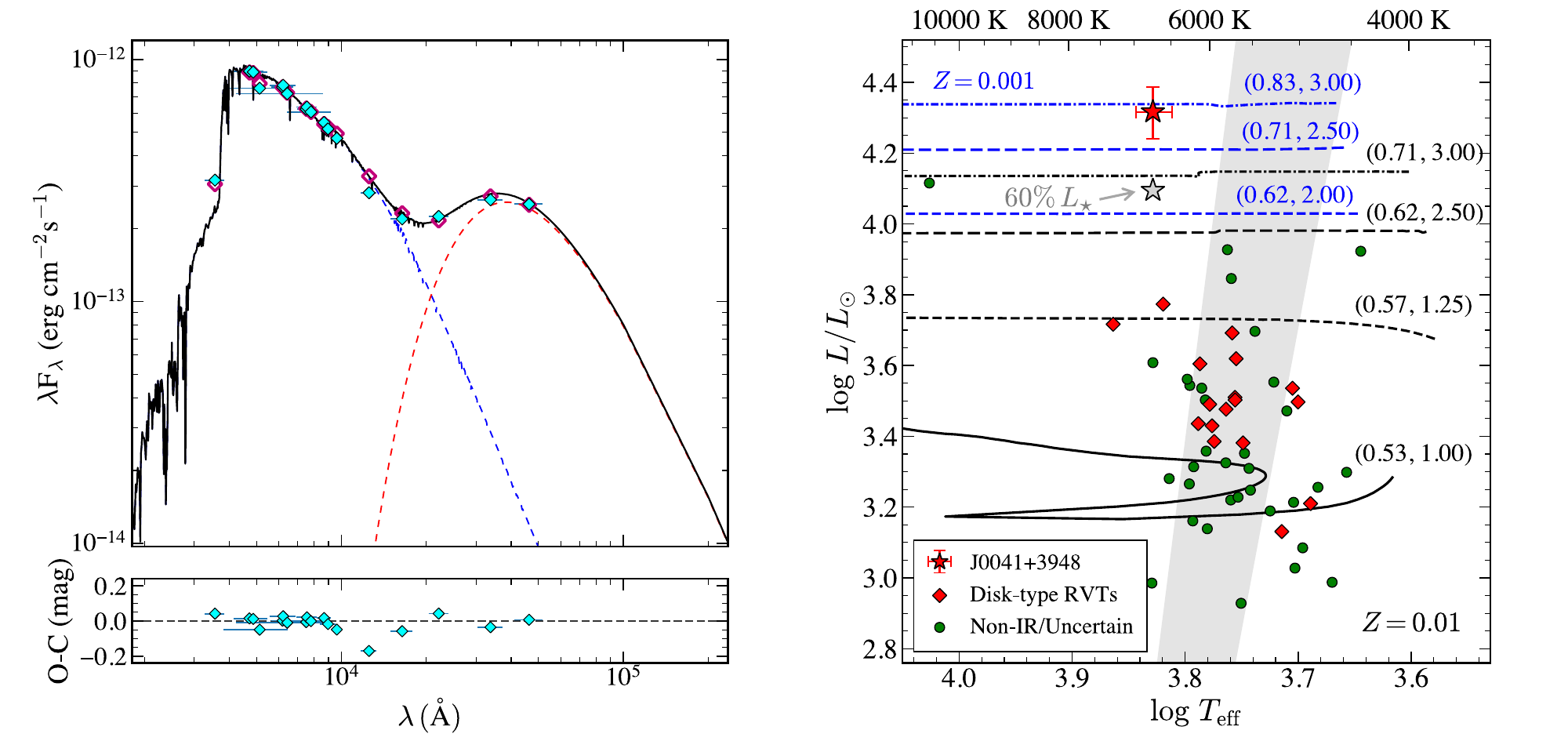}
\caption{Left: SED of J0041+3948. Cyan diamonds show the observed photometry and magenta diamonds denote the synthetic photometry. The solid black curve is the best-fitting model, decomposed into a stellar component (blue dashed) and a dust component (red dashed). 
Right: HR diagram showing the position of J0041+3948 (red asterisk with error bars). The gray asterisk marks a reference luminosity of $0.6\,L_\star$. Red diamonds represent disk-type RVTs in the Magellanic Clouds from \citet{Manick2018}, and green circles denote RVTs without IR excess or with uncertain classification. Also shown are post-AGB evolutionary tracks from \citet{Miller2016} at $Z=0.01$ (black curves with various line styles) and at $Z=0.001$ (blue curves), with the corresponding final and initial masses labeled above the tracks. The gray shaded region indicates the boundaries of the instability strip from \citet{Desomma2021} for $Z=0.004$.
\label{fig: sed}}
\end{figure*}

In the right panel of Figure\,\ref{fig: location}, we compare the velocity of J0041+3948 in position-velocity space with that of the GSS, traced by DESI stars from \citet{Dey2023}. Heliocentric velocities were converted to the Galactic Standard of Rest (GSR) using a Solar velocity of [12.9, 245.6, 7.78]\,$\kms$ in [$x, ~y, ~z$] and subtracting the systemic velocity of M31 \citep[$-113.7\,\kms$;][]{McConnachie2012}. Because J0041+3948 lies at the edge of the GSS region defined by \citet{Dey2023}, we adopted a slightly broader sample with position angles between 140\arcdeg\ and 190\arcdeg, including only objects located outside the optical disk ellipse to avoid contamination. As illustrated, the bulk of M31 halo stars form a smoother distribution in the figure, whereas the GSS stands out as a tight, blueshifted sequence. $\Delta V_{\text{GSR}}$ ranges from $\sim -300\,\kms$ at $R_{\rm proj}=0.5^\circ$ to $\sim -50\,\kms$ at $R_{\rm proj}=4^\circ$,  and the sequence is markedly steeper at smaller $R_{\rm proj}$. With a measured radial velocity of $\sim -511\,\kms$ (see Appendix~\ref{sec:spec_log}), corresponding to a $\Delta V_{\text{GSR}}=-212.6\,\kms$, J0041+3948 clearly follows this pattern, and the combined spatial and kinematic evidence indicates that it is likely associated with the GSS.

\section{Physical Properties} \label{sec:physical properties}

\subsection{Spectral Energy Distribution}\label{subsec:sed}

In the left panel of Figure\,\ref{fig: sed} the SED of J0041+3948 is shown. It exhibits the typical disk-type morphology, which can be decomposed into two components: a stellar continuum in the optical and a clear near-IR excess. To derive physical properties, we fitted the SED using a stellar atmosphere model from \citet{ck03} at [Fe/H]$=-0.5$ plus a single-temperature blackbody. The fitting was performed with the Python package \texttt{SPEEDYFIT} \citep{Vos2017, Vos2018}, which employs a Markov chain Monte Carlo (MCMC) approach to model both components simultaneously. We adopted the extinction law of \citet{Cardelli1989} with $R_V = 3.1$. We ran the MCMC with 25 walkers and 10,000 steps, adopting Gaussian priors on the distance to the GSS based on \citet{Conn2016} and on $T_{\text{eff}}$ and $\log g$ based on our spectroscopic estimates, and uniform priors on the remaining parameters.
The fitted model reproduces the observed fluxes reasonably well, with typical residuals within $\sim$0.05\,mag. Pulsation amplitudes reach  $\sim$0.1\,mag in the blue optical bands and are smaller at longer wavelengths, introducing additional scatter in the compiled SED. However,  this effect is partly mitigated because some of the input photometry is averaged over multiple epochs. To account for potential systematics, we adopt $2\sigma$ confidence intervals for the fitted parameters. The derived values are summarized in Table~\ref{tab: Properties}.

The fit yields an effective temperature of $6738$\,K, a stellar radius of $106^{+8}_{-7}\,R_\odot$, and a luminosity of $10^{4.32}\,L_{\odot}$ for the stellar component, thereby establishing the post-AGB nature of J0041+3948. A blackbody component at $\sim1000$\,K indicates hot dust near the sublimation temperature. Sustaining such hot emission requires dust to remain near the star in a gravitationally bound reservoir, since at a typical AGB wind speed of $\sim10\,\kms$ \citep[e.g.,][]{Hofner2018}, freely expanding material would reach cooler radii within a few years. In post-AGB stars, this emission is now well established to arise from a stable, compact dusty disk and is a robust indicator of binarity, since circumbinary disks form through interaction with a companion. By contrast, a single post-AGB star would drive a freely expanding wind with cooling dust whose emission would peak in the mid-IR \citep{Gezer2015}. We therefore infer that J0041+3948 is a binary system.

It is worthy of notice that SED-derived parameters for post-AGB binaries should be treated with caution, as circumstellar scattered light can bias the results. The effect is most severe in systems with $L_{\rm dust}/L_\star>1$, where the post-AGB star may be seen predominantly in scattered light. Unfortunately, a quantitative correction is not straightforward and depends on several coupled factors, including the disk geometry, inclination, and the kind of dust grains. For J0041+3948, the SED fit yields $L_{\rm dust}/L_\star\approx0.32$, a typical value for disk sources, so we conclude that scattered light does not make a major contribution to the observed SED. Nevertheless, in the discussion that follows we adopt a scattered-light fraction of 40\% as a conservative upper limit to test the robustness of our conclusions, guided by the interferometric results of \citet{Hillen2013}. In addition, we do not correct for flux from the unseen secondary, consistent with the companion-mass distribution for post-AGB binaries \citep{Oomen2018}, which implies low-mass main-sequence companions with negligible optical luminosities.

\subsection{HR Diagram}\label{subsec:HRD}

Armed with the temperature and luminosity derived from the SED fit, we place J0041+3948 on the HR diagram in the right panel of Figure~\ref{fig: sed}, and compare its position with the theoretical post-AGB evolutionary tracks of \citet{Miller2016}. Because the metallicity of J0041+3948 is unknown, we adopt tracks with $Z=0.001$ and $Z=0.01$ to bracket the plausible range for the GSS and inner halo at this location, with $Z=0.01$ likely being more representative \citep{Conn2016, Dey2023}. In these models, the quoted final mass is the white dwarf (WD) mass and is essentially equal to the current stellar mass for evolved post-AGB stars, while the initial mass refers to the progenitor mass. Although the tracks were computed for single stars and do not include binary effects, we assume that interactions with the companion and the circumbinary disk do not substantially alter the core mass-luminosity relation. As shown in the figure, J0041+3948 occupies the high-luminosity yellow post-AGB regime in the HR diagram.

Turning to quantitative constraints, J0041+3948's locus is most consistent with a final mass of $0.83\,M_\odot$ at $Z=0.001$, with the lower bound extending to about $0.71\,M_\odot$ when uncertainties are included. Assuming $Z=0.01$, its luminosity exceeds the upper envelope of the available tracks, but this does not materially affect the estimate because the final mass shows only a mild dependence on metallicity. Figure~\ref{fig: sed} also shows a reference point at $0.6\,L_\star$, corresponding to a conservative 40\% scattered-light contribution. At this luminosity, the source falls between the tracks for final masses of $0.62\,M_\odot$ and $0.71\,M_\odot$. Therefore, the final mass likely lies in the range $0.62-0.83\,M_\odot$. 

Mapping the final mass to a progenitor mass is inherently uncertain. According to the evolution models from \citet{Miller2016}, a final mass of $0.71-0.83\,M_\odot$ corresponds to an initial mass of $2.5-3.0\,M_\odot$ at $Z=0.001$. At $Z=0.01$, the same mass range implies an initial mass larger than $3.0\,M_\odot$. We adopt $4.0\,M_{\odot}$ as a conservative upper bound, based on the empirical initial-final mass relation from \citet{Cummings2018}, which exhibits a break associated with the onset of second dredge-up in this mass range. Allowing for a scattered-light contribution up to 40\%, the lower bound on the final mass ($0.62-0.71\,M_\odot$) would imply progenitors of roughly $2.0-3.0\,M_{\odot}$. Taken together, we adopt an initial mass estimate of $M_{\rm i}\approx 2.0-4.0\,M_{\odot}$, with the dominant uncertainties arising from metallicity and the fraction of scattered light.

\begin{figure*}
\centering 
\includegraphics[width=0.95\textwidth]{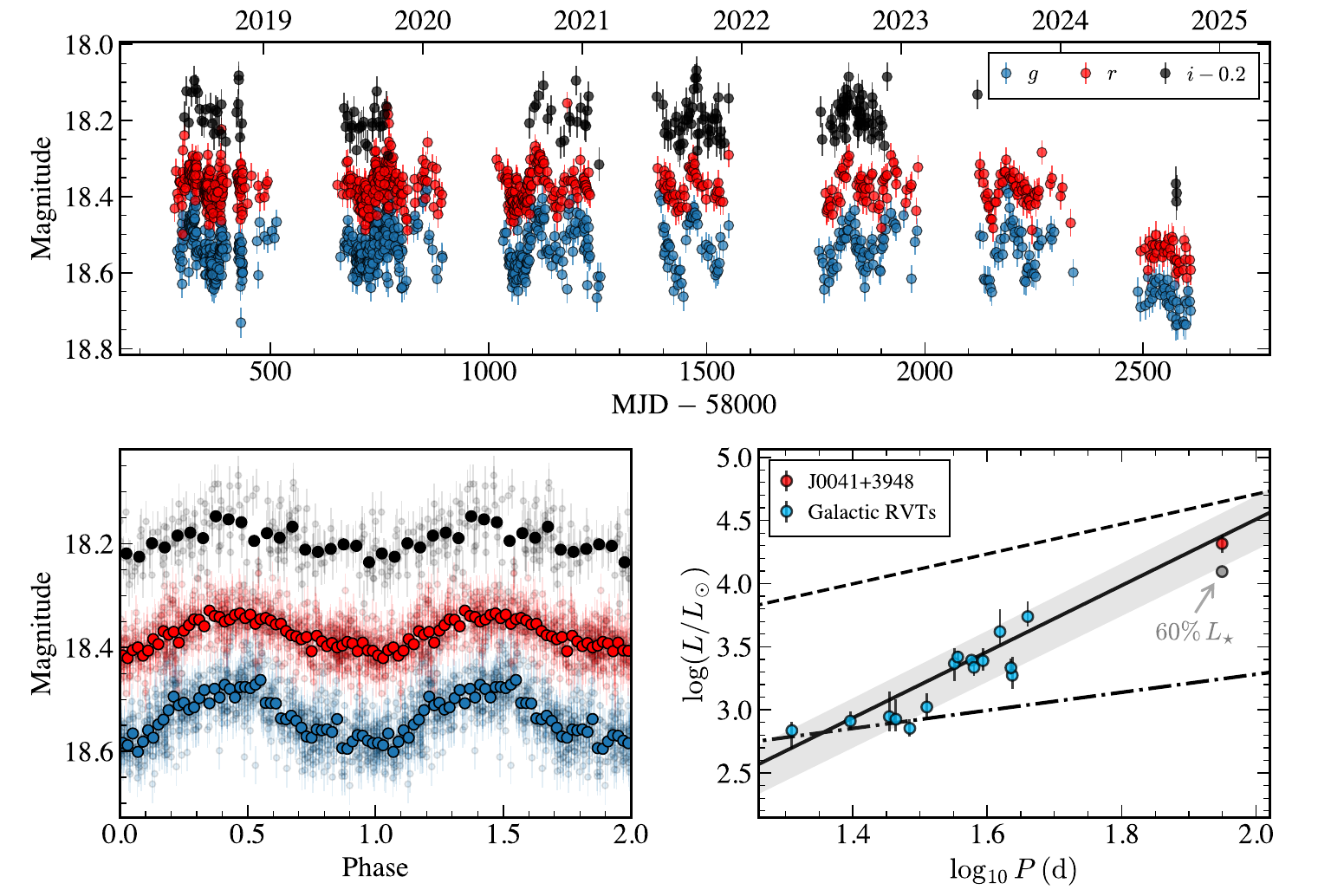}
\caption{Top: ZTF $gri$ light curves with magnitudes offset vertically for clarity.
Bottom left: Phase-folded light curves ($\mathrm{MJD}<60400$) after removing outliers. The binned circles show mean values in phase bins of 0.02 for $g$ and $r$ and 0.05 for $i$. 
Bottom right: PLR for TIICs. J0041+3948 is shown as a red circle with error bars, and the gray circle marks a reference luminosity of $0.6\,L_\star$. The solid black line and its shaded $1\sigma$ interval show the PLR derived by \citet{Bodi2019} for Galactic long-period RVTs. The dashed line shows the PLR for LMC classical Cepheids from \citet{Groenewegen2023}, and the dotted line shows the PLR for Magellanic TIICs from \citet{Groenewegen2017}.
\label{fig: lc}}
\end{figure*}

\subsection{Photometric Variability}\label{subsec:lc}
The top panel of Figure\,\ref{fig: lc} displays the ZTF light curves of J0041+3948. To remove poor-quality detections, we applied the following quality cuts: \texttt{catflags}$=0$ (photometric/image quality), $\lvert\texttt{sharp}\rvert<0.25$ and \texttt{chi}$<2$ (PSF-fit metrics), and \texttt{airmass}$<1.5$. The source exhibits a broadly stable mean brightness with intrinsic variability, followed by a dimming that occurred in the first half of 2024. We searched for periodicity with the \texttt{Astropy} \citep{astropy1, astropy2, astropy3} Lomb-Scargle implementation \citep{Lomb1976, Scargle1982, VanderPlas2018}. After removing outliers and restricting to data prior to the dimming ($\mathrm{MJD}<60400$), we determined a strong period of 88.96\,days from the $r$-band light curve. The bottom left panel of Figure~\ref{fig: lc} shows the $g$, $r$, and $i$ light curves folded at this period.

The long pulsation period and post-AGB nature place J0041+3948 in the realm of RVTs. In the right panel of Figure~\ref{fig: sed}, we compare its HR-diagram position with Magellanic RVTs from \citet{Manick2018}, who showed that dusty RVTs are, on average, more luminous than non-dusty ones and possibly descend from more massive progenitors. J0041+3948 lies near the instability strip and appears to be a higher-luminosity extension of the known RVTs. Its low pulsation amplitude is consistent with a more evolved status relative to typical post-AGB RVTs \citep{Kiss2007}, a property also seen in several objects from \citet{Manick2018}. 

The period-luminosity relation (PLR) for RVTs within the TIIC class remains less well constrained. Previous studies indicate that these long-period variables follow a steeper PLR than their shorter-period, less luminous counterparts \citep[e.g.][]{McNamara1995, Groenewegen2017, Bodi2019}, and this trend is pronounced in the bottom right panel of Figure~\ref{fig: lc}. The PLR of Galactic long-period RVTs from \citet{Bodi2019} is systematically steeper than the PLR obtained for the full TIIC population \citep{Groenewegen2017} and approaches classical-Cepheid luminosities by $\sim 100$\,d. Extrapolating the PLR from \citet{Bodi2019} predicts a luminosity in close agreement with our derived value for J0041+3948. Moreover, even after accounting for a 40\% scattered-light fraction, its luminosity remains close to the $1\sigma$ confidence interval of the PLR in Figure~\ref{fig: lc}. Therefore, we classify J0041+3948 as a long-period TIIC and a high-luminosity analogue of the known dusty RVTs. Indeed, it is the most massive and luminous post-AGB TIIC yet found.

\begin{figure*}
\centering 
\includegraphics[width=1.0\textwidth]{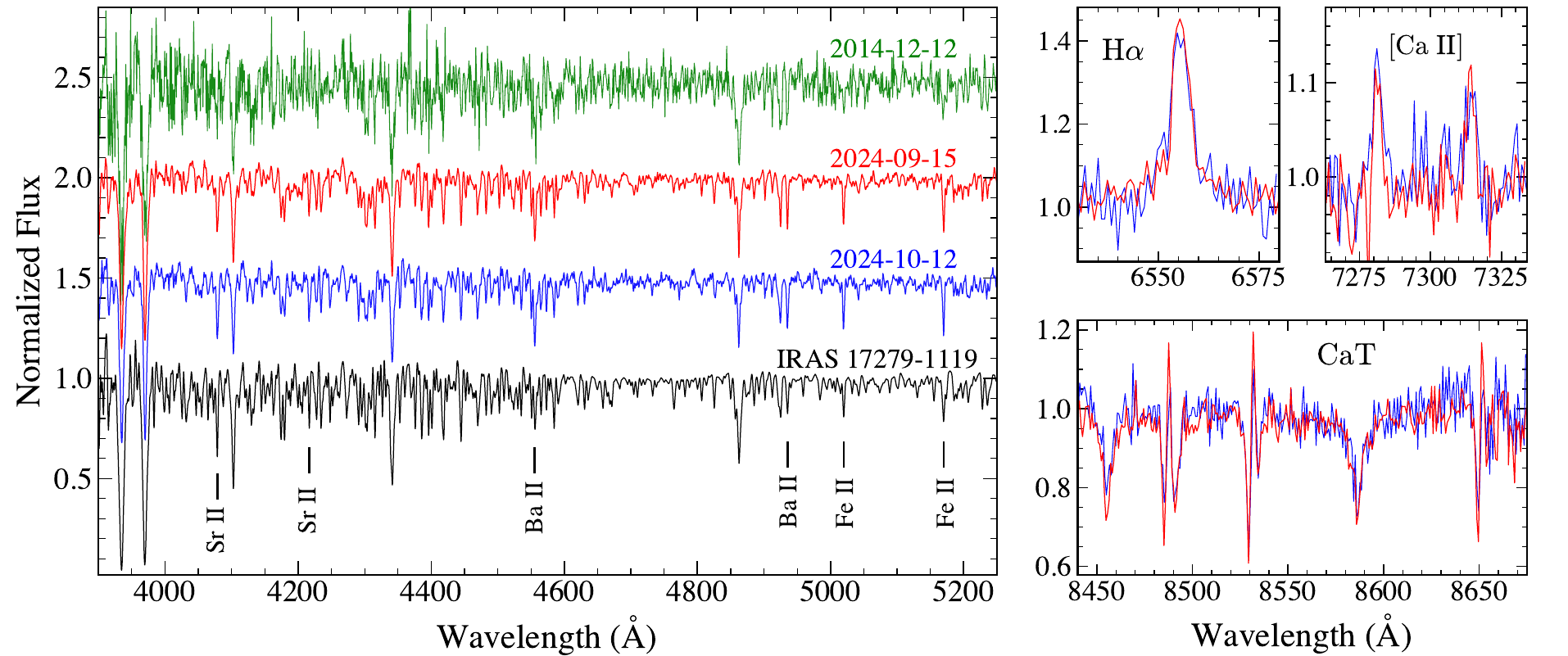}
\caption{Left: Normalized spectra of J0041+3948 at three epochs (top three), together with a template spectrum of IRAS 17279$-$1119, which is a a known $s$-process-enriched post-AGB binary. All spectra are degraded to the LAMOST resolution for clarity and shifted to the rest frame. Key spectral features are labeled.
Right: Zoomed regions in the two DBSP spectra around H$\alpha$, [\ion{Ca}{2}] $\lambda\lambda 7291,7324$, and CaT, shown at their native resolution and in the observed frame.
\label{fig: spec}}
\end{figure*}

The pulsation variability of J0041+3948 persists in the dimmed state. Using the median magnitude in each observing season, we found that the source faded by $\sim0.14$\,mag in the $g$ band and $\sim0.17$\,mag in the $r$ band. Although the uncertainties are relatively large, the object became systematically bluer as it faded. This behavior supports our earlier caution that a non-negligible scattered-light component should be considered, while extinction alone would be expected to produce either reddening or little change in color. Given that the source remained photometrically stable for $\sim2000$ days prior to the dimming, and that the Gaia DR3 light curves (not shown) also exhibit stable brightness from 2014 August to 2017 May, it is unlikely that the dimming was caused by disk occultation, which would be expected to recur periodically. The system is therefore likely viewed at low inclination, and the event may instead reflect a dusty circumstellar environment in which opaque dust clouds move more stochastically within the system. As these clouds move across the line of sight, the extinction increases and the fraction of scattered light rises. At the same time, the motion of these clouds may redirect additional scattered light into the line of sight, producing a bluer color at lower flux. Continued monitoring will be essential to constrain the nature of this event.

\subsection{Spectral Features}\label{subsec:spec}
We display the three normalized spectra of J0041+3948 in Figure\,\ref{fig: spec}. In the blue (left panel), the spectra are characterized by numerous strong, narrow absorption lines from singly ionized metal species. Notably, we detect the resonance \ion{Ba}{2} 4554 and 4934\,\AA\ lines, which remain prominent even at our modest resolution and point to $s$-process enrichment. We also identify features of other $s$-process and refractory elements, including but not limited to \ion{Sr}{2}, \ion{Ti}{2}, and \ion{Fe}{2}. Across the three spectra, no evident changes in spectral features or radial velocities are found (see Appendix~\ref{sec:spec_log} for details). Given the binary nature inferred from the SED, J0041+3948 appears to be another chemically peculiar post-AGB disk source (also a TIIC variable) that shows $s$-process enhancement. This is the first such example identified beyond the Milky Way and the Magellanic Clouds. For comparison, we included in the figure the FEROS spectrum of IRAS 17279$-$1119, which is also a $s$-process enriched post-AGB binary \citep{DeSmedt2016}. The reduced spectrum was retrieved from the ESO Phase 3 archive \citep{eso}. 

In contrast to the common depletion pattern in post-AGB binaries, these few outliers demonstrate that a circumbinary disk is necessary for depletion but not sufficient. \citet{Menon2024} analyzed three $s$-process enriched post-AGB binaries and ruled out both extrinsic pollution and inherited galactic-scale enrichment. We disfavor an extrinsic origin involving a WD companion, as in classical barium stars, because producing such a massive post-AGB star would require the progenitor to accrete an implausibly large amount of mass from the companion (see Section~\ref{sec:formation pathway}). An inherited explanation is also unlikely for a source in M31's halo. Therefore, the observed enrichment in J0041+3948 is most likely intrinsic. It is possible that the dust-gas separation process failed in these objects, causing the re-accreted material to remain rich in refractory elements. A quantitative abundance analysis based on high-resolution spectroscopy will be valuable to better understand such abnormal phenomenon.

No features are identified in the red LAMOST spectrum because of the low S/N. In the red arm of the two DBSP spectra (right panels of Figure\,\ref{fig: spec}), we detect H$\alpha$ in emission with broad wings, forbidden [\ion{Ca}{2}] $\lambda\lambda 7291,7324$ emission, and \ion{Ca}{2} near-IR triplet (CaT) emission superposed on photospheric absorption, signaling the presence of stellar winds and circumstellar material. Among these, the [\ion{Ca}{2}] lines are rarely observed in post-AGB stars \citep{Klochkova2020}. \citet{Aret2012} showed that these forbidden lines are excellent tracers of the dense regions within the circumstellar disks of B[e] supergiants. By contrast, circumbinary disks around post-AGB binaries are generally cooler and predominantly dusty, where calcium is expected to be strongly depleted from the gas phase. Therefore, the detection of these lines points to an additional dense, hot gaseous component, likely the accretion disk around the companion. We measured a radial velocity of $\sim -490\pm10\,\kms$ for the [\ion{Ca}{2}] lines, which is slightly redshifted relative to the stellar velocity, corresponding to a $\sim2\sigma$ discrepancy. This may support a distinct origin, although the measured velocity is uncertain due to telluric contamination.

\section{Possible Formation Pathway}\label{sec:formation pathway}

The origin of J0041+3948 remains puzzling in the context of the GSS, widely interpreted as the remnant of a galaxy merger. Deep HST pencil-beam imaging revealed that GSS region hosts a composite stellar population with an extended star formation history (SFH), including intermediate-age components ($\sim4-11$\,Gyr) and a relatively small yet notable fraction of $\sim$2\,Gyr-old stars \citep{Brown2006, Bernard2015}. While the older population is generally interpreted as tracing the main stellar component of the disrupted satellite galaxy, recent simulations suggest that the younger population could either be stars originally belonging to the M31 disk, or young stars formed from gas (primarily in the rebuilt thin disk) during the galaxy interaction and later displaced to the present location \citep{Hammer2018, Tsakonas2025}. Using the main-sequence plus RGB lifetimes from \citet{Miller2016}, this age distribution translates to initial progenitor masses of roughly $1.0$ to $1.6\,M_\odot$, although the progenitor-mass estimates for the GSS population remain approximate due to its extended SFH. By contrast, the inferred progenitor mass of J0041+3948 is $\sim 2.0-4.0\,M_\odot$ (see Section~\ref{subsec:HRD}), indicating that it is significantly more massive and younger than typical GSS stars. We further find that it is substantially more luminous than any other TIICs identified in this region by \citet{Kodric2018}. Together, these facts raise the question of how such a star could have formed.

Here we propose a possible formation pathway via stellar merger that qualitatively accounts for the observed properties of J0041+3948 and satisfies the evolutionary constraints imposed by the GSS formation history. In detail, we consider an initially hierarchical triple system, dynamically stable by construction, in which two $\sim1.0-1.6\,M_\odot$ stars form a close inner binary and a low-mass tertiary orbits them on a much wider outer orbit. As one of the inner components evolves off the main sequence toward the RGB, close binary interactions, enhanced by the eccentric Kozai-Lidov (EKL) mechanism \citetext{\citealp{Kozai1962, Lidov1962}; see \citealp{Naoz2016} for a review} induced by the tertiary, shrink the inner orbit and drive an efficient merger that produces a $\sim2.0-3.2\,M_\odot$ remnant. This remnant then evolves rapidly to the AGB, where strong mass loss, together with interaction with the tertiary, leads to the formation of a circumbinary disk. After losing most of its envelope, the remnant is now crossing the instability strip and pulsates as an extremely luminous, long-period TIIC.

In this scenario, the present post-AGB star is the descendant of the merged inner binary, while the unseen companion is the original tertiary. Given the uncertainties in translating the GSS SFH into progenitor masses, the merger of two typical GSS stars can plausibly produce a remnant of up to $\sim3.2\,M_\odot$. This is broadly consistent with our inferred $2.0-4.0\,M_\odot$ progenitor-mass range for J0041+3948.

Hierarchical triples are observationally common \citep[e.g.][]{Raghavan2010, Duchene2013, DeRosa2014, Shariat2025b}. Furthermore, a growing body of observational and theoretical work also shows that EKL cycles can efficiently drive inner-binary mergers in hierarchical triples \citep[e.g.][]{Naoz2014, Toonen2016, Toonen2020, Shariat2023, Kummer2025, Shariat2025a, Perets2025}, providing channels for forming exotic binaries such as blue straggler stars \citep[BSSs;][]{Geller2013, Leiner2025} and barium stars \citep{Gao2023}. Although our current constraints on the orbital configuration of J0041+3948 remain limited, the pathway that we propose is a natural outcome of stellar multiplicity.
In this channel, the inner-binary merger can produce a BSS-like binary system with a significantly rejuvenated primary, which subsequently evolves to the post-AGB phase. J0041+3948 may therefore offer a glimpse of the subsequent evolution of some BSSs formed through triple-driven mergers. Given the $s$-process enrichment and the evidence for an accreting companion, the post-AGB primary likely transfers enriched material to the companion. In the future, J0041+3948 will likely become a barium star system, thereby providing a plausible pathway to the observed barium stars with high-mass WD companions \citep{Escorza2023}.

For completeness, we consider whether J0041+3948 could be a stable mass transfer product in a binary that now hosts a WD companion, analogous to binary barium stars and some BSSs. To match both the observed $s$-process enrichment and the inferred progenitor mass, a typical GSS star would need to accrete at least $\sim0.4-3.0\,M_\odot$ from an evolved AGB donor, far exceeding the accreted masses typically considered for these chemically polluted systems \citep[$\lesssim 0.5\,M_\odot$; e.g.][]{Stancliffe2021}. Moreover, our spectra show no evidence for symbiotic activities, which would be expected if the companion were now an accreting compact object. We therefore do not pursue this scenario further, although future high-resolution abundance analyses will provide a more stringent test of this possibility.

\section{Summary}\label{sec:summary}
In this Letter, we report the discovery of LAMOST J0041+3948, the most luminous known post-AGB TIIC, located in the outskirts of M31 where the GSS crosses the inner halo. The source exhibits a disk-type SED with a broad near-IR excess, interpreted as emission from hot dust in a circumbinary disk, implying a binary system. An SED fit yields an effective temperature of $6738_{-262}^{+234}$\,K and luminosity of $\log L/L_{\odot}=4.32_{-0.08}^{+0.07}$ for the post-AGB star, together with a $\sim1000$\,K dust component near the sublimation temperature. Comparison with evolutionary tracks suggests that this luminosity is best matched with a stellar mass of $\sim0.83\,M_{\odot}$, mapping to a $\sim3.0-4.0\,M_{\odot}$ progenitor. Accounting for a fraction of scattered-light contribution of up to $40\%$ would relax the initial mass to $\sim2.0-4.0\,M_{\odot}$. ZTF light curves reveal a pulsation period around $89$\,d that lies on the PL relation for long-period RVTs. It recently underwent a dimming event during which the source faded and became slightly bluer. This abnormal behavior supports the inclusion of a non-negligible scattered-light component and points to a dusty circumstellar environment in this binary system. Despite the presence of a circumbinary disk, the photospheric depletion pattern commonly seen in post-AGB binaries is absent. Instead, the observed spectra show numerous absorption lines of refractory species, including strong \ion{Ba}{2} features that points to $s$-process enrichment, which we argue is most likely intrinsic. Moreover, we find spectroscopic clues for an accretion disk around the companion.

Intriguingly, although the position and radial velocity of J0041+3948 are consistent with those of the GSS, it is significantly younger and more massive than both the stellar population expected in the stream and other TIICs in the vicinity. Based on the available observations, we propose a  formation scenario through a stellar merger, in which the observed post-AGB star is the merger product of two $\sim1.0-1.6\,M_\odot$ stars, and the present binary system descends from a hierarchical triple system. At an earlier evolutionary stage, the system was displaced to its current location by tidal disruption or perturbations associated with the recent galaxy merger.

J0041+3948 is a rare find. It illustrates how binarity and its interaction with a circumbinary disk can shape the observable properties of evolved stars. While its long-period pulsation marks it as an extremely luminous TIIC, it underscores the importance and complexity of multiplicity in stellar evolution. Moreover, it is a firsthand witness of galaxy interaction. Future observations, both photometric and spectroscopic, are encouraged to determine key properties such as metallicity and orbital elements, which will further clarify the nature of this object. In addition, we note that multiplicity and merger events have been proposed to explain several peculiar classical Cepheids in binary systems \citep{Pilecki2022}, and the progenitor masses of those Cepheids overlap with our estimate for J0041+3948 (which is a TIIC). It would be interesting to explore the evolutionary links and distinctions between these systems.

\section*{Acknowledgments}
We are grateful to the anonymous referee, whose excellent comments and suggestions significantly improved this article. This work is partially supported by the National Natural Science Foundation of China 12173034, 12322304, 12173047, 12322306, and 12373028, the National Natural Science Foundation of Yunnan Province 202301AV070002 and the Xingdian talent support program of Yunnan Province. We acknowledge the science research grants from the China Manned Space Project with no. CMS-CSST-2025-A11. X.C. and S.W. acknowledge support from the Youth Innovation Promotion Association of the Chinese Academy of Sciences (Nos. 2022055 and 2023065).

Guoshoujing Telescope (the Large Sky Area Multi-Object Fiber Spectroscopic Telescope LAMOST) is a National Major Scientific Project built by the Chinese Academy of Sciences. Funding for the project has been provided by the National Development and Reform Commission. LAMOST is operated and managed by the National Astronomical Observatories, Chinese Academy of Sciences.

This research uses data obtained through the Telescope Access Program (TAP), which is funded by the National Astronomical Observatories, Chinese Academy of Sciences, and the Special Fund for Astronomy from the Ministry of Finance. Observations obtained with the Hale Telescope at Palomar Observatory were obtained as part of an agreement between the National Astronomical Observatories, Chinese Academy of Sciences, and the California Institute of Technology. We are grateful to the staffs of Palomar Observatory for assistance with the observations and data management.

This work is based on observations obtained with the Samuel Oschin 48-inch Telescope at the Palomar Observatory as part of the Zwicky Transient Facility (ZTF) project.  ZTF is supported by the National Science Foundation under Grants No. AST-1440341 and AST-2034437 and a collaboration including current partners Caltech, IPAC, the Oskar Klein Center at Stockholm University, the University of Maryland, University of California, Berkeley, the University of Wisconsin at Milwaukee, University of Warwick, Ruhr University, Cornell University, Northwestern University and D`rexel University. Operations are conducted by COO, IPAC, and UW. This work has made use of data from SDSS, PS1, Gaia, UKIRT, and WISE. Based on data obtained from the ESO Science Archive Facility with DOI: \url{https://doi.eso.org/10.18727/archive/24}.

\software{\texttt{PypeIt} \citep{pypeit1, pypeit2}, \texttt{SPEEDYFIT} \citep{Vos2017, Vos2018},
\texttt{Numpy} \citep{numpy}, \texttt{Matplotlib} \citep{Matplotlib}, \texttt{Astropy}\citep{astropy1,astropy2,astropy3}, \texttt{Pandas} \citep{pandas}, \texttt{Scipy} \citep{scipy}.}

\appendix

\section{Spectroscopic Observing Log  and Radial Velocity Measurements}\label{sec:spec_log}
We provide in Table~\ref{tab: observations} a record of  spectroscopic observations in chronological order.

Radial velocities were measured from three spectra following \citet{Chen2025b}. We used the high-resolution ($R\sim 48000$) FEROS spectrum of IRAS 17279$-$1119 as the template because its close spectral match. Using lines of singly ionized metals in pure absorption, primarily \ion{Fe}{2}, \ion{Ti}{2}, and \ion{Ba}{2}, we measured a velocity of $56.6\pm0.4\,\kms$ for the template. We degraded the template to match the spectral resolutions of LAMOST and DBSP, respectively, and measured the velocities using the cross-correlation function (CCF) defined by \citet{Zhang2021}. The CCF was computed over $4000-5350$\,\AA, and the velocity at the CCF maximum was obtained by maximizing the function with \texttt{scipy.optimize.minimize} using the Nelder-Mead algorithm \citep{Nelder1965}. To estimate uncertainties, we performed 500 Monte Carlo simulations by injecting Gaussian noise into the observed spectra. Because the method estimates error based solely on the photon noise, it tends to underestimate the errors. Therefore, we adopted 2$\sigma$ confidence intervals to account for potential systematics. The resulting values are $-490.8\pm10.9\,\kms$ for the LAMOST spectrum, and $-511.2\pm1.6\,\kms$ and $-510.3\pm2.7\,\kms$ for the first and second DBSP spectra, respectively.

The LAMOST spectrum has low S/N, leaving only a few strong Balmer lines available for a reliable velocity determination. However, these lines may not reflect the stellar systemic velocity because they often show non-photospheric contributions from stellar winds or circumbinary material, so we do not use this measurement. Within the uncertainties, we detect no radial-velocity variation over the $\sim1$\,month separation between the two DBSP epochs. This is not unexpected, since post-AGB binaries often have orbital periods of a few hundred days to several years \citep{Oomen2018}. A low inclination angle (see Section~\ref{subsec:lc}) would further reduce the radial velocity amplitude. Finally, we adopt the velocity from the highest-S/N DBSP spectrum in Table~\ref{tab: Properties} and note that minor velocity offsets from undetected orbital motion would not affect our conclusions or the association with the GSS.

\begin{table*}
\caption{Summary of Spectroscopic Observations\label{tab: observations}}
\setlength{\tabcolsep}{15.pt}
\begin{center}
\begin{tabular}{cccccccc}
\hline
\hline
Date  & Telescope/ & Wavelength   & FWHM  & Exposure & Seeing \\
 (UT)  & Instrument & Range (\AA) & (\AA) &  (s) & (\arcsec) & \\

\hline
2014-12-12 & LAMOST LRS & 3700--9085 & 2.5 & 3$\times$1800 & 3.1\\
2024-09-15 & P200/DBSP-B1200 & 3850--5395 & 2.11 & 6$\times$1800 & 1.2\\
           & P200/DBSP-R600  & 5740--9100 & 4.19 & 7$\times$1500 & 1.2\\
2024-10-12 & P200/DBSP-B1200 & 3780--5325 & 2.11 & 4$\times$1800 & 1.5\\
           & P200/DBSP-R600  & 5700--9055 & 4.19 & 4$\times$1800 & 1.5\\
\hline
\end{tabular}
\end{center}
\end{table*}

\bibliography{sample701.bib}{}
\bibliographystyle{aasjournalv7}

\end{document}